# Deep learning to generate *in silico* chemical property libraries and candidate molecules for small molecule identification in complex samples

Sean M. Colby, Jamie R. Nuñez, Nathan O. Hodas, Courtney D. Corley, Ryan R. Renslow*

Pacific Northwest National Laboratory, Richland, WA, USA.

* ryan.renslow@pnnl.gov

**ABSTRACT:** Comprehensive and unambiguous identification of small molecules in complex samples will revolutionize our understanding of the role of metabolites in biological systems. Existing and emerging technologies have enabled measurement of chemical properties of molecules in complex mixtures and, in concert, are sensitive enough to resolve even stereoisomers. Despite these experimental advances, small molecule identification is inhibited by (i) chemical reference libraries (e.g. mass spectra, collision cross section, and other measurable property libraries) representing <1% of known molecules, limiting the number of possible identifications, and (ii) the lack of a method to generate candidate matches directly from experimental features (i.e. without a library). To this end, we developed a variational autoencoder (VAE) to learn a continuous numerical, or latent, representation of molecular structure to expand reference libraries for small molecule identification. We extended the VAE to include a chemical property decoder, trained as a multitask network, in order to shape the latent representation such that it assembles according to desired chemical properties. The approach is unique in its application to metabolomics and small molecule identification, with its focus on properties that can be obtained from experimental measurements (*m/z*, CCS) paired with its training paradigm, which involved a cascade of transfer learning iterations. First, molecular representation is learned from a large dataset of structures with *m/z* labels. Next, *in silico* property values are used to continue training, as experimental property data is limited. Finally, the network is further refined by being trained with the experimental data. This allows the network to learn as much as possible at each stage, enabling success with progressively smaller datasets without overfitting. Once trained, the network can be used to predict chemical properties directly from structure, as well as generate candidate structures with desired chemical properties. Our approach is orders of magnitude faster than first-principles simulation for CCS property prediction. Additionally, the ability to generate novel molecules along manifolds, defined by chemical property analogues, positions DarkChem as highly useful in a number of application areas, including metabolomics and small molecule identification, drug discovery and design, chemical forensics, and beyond.

## INTRODUCTION

High throughput small molecule identification in complex samples typically requires the comparison of experimental features (e.g., *m/z*, chromatographic retention times) to corresponding reference values in libraries in order to build evidence for the presence of a particular molecule. Libraries can be determined experimentally through analysis of authentic reference materials (i.e., standards), or through *in silico* calculation of chemical properties and prediction of analytical features[1-6]. The former is preferred[7-8], and is currently the gold standard approach for library-building, primarily due to the assumed lower associated variance for properties derived from modern analytical platforms, and thus higher levels of assigned confidence to identifications. However, most compounds are not available for purchase as authentic reference material, cannot be isolated or easily synthesized, or are simply yet unknown[9-11]. In addition, the experimental route for library building is costly and time consuming. In contrast, *in silico* methods can yield reference values rapidly, facilitating the creation of much larger libraries than reasonably achievable through experimental methods.

*In silico* library-building methods for applications in metabolomics vary, ranging from first-principles physics simulations[4, 6, 12], to property-based machine learning approaches[1-2, 13-16]. While useful, both methods have limitations: first-principles approaches can require a deep understanding of the underlying physics, which may not be well understood, and substantial compute time to yield accurate predictions. Furthermore, it is currently infeasible to use first-principles-based methods in a generative manner, i.e., to directly create molecular structures with desired properties. Conversely, machine learning approaches generally require large training sets and predictions are typically constrained to molecules similar to those found within the training set. Thus, machine learning approaches may not necessarily generalize to novel molecules outside of the chemical classes represented by the training set.

Recent interest in chemical structure-based deep learning approaches have shown promise[17-24], particularly in the application of variational autoencoders (VAEs) [25] and other generative approaches for learning a continuous numerical, or latent, representation of molecular structure[17, 20-21, 23]. These networks take SMILES (simplified molecular line entry system) strings as input and, in a semi-supervised configuration, predict the same sequence of characters as output, after perturbation by noise. Importantly, recent works have begun coupling the latent representation of molecular structure to property predictor subnetworks, such as lipophobicity (logP), quantitative estimate of drug-likeness (QED), synthetic accessibility score (SAS), lowest unoccupied molecular orbital (LUMO), and the electronic spatial extent (r2)[23]. This yields latent space entanglement, wherein the vectors describing molecular inputs begin to encode both structure and property, with implications in "molecular optimization." That is, traversing latent space to generate molecules with desired properties. However, these deep learning applications have been largely limited to drug design[17, 21, 23] and other industrial pursuits[20, 22, 26], which has been reflected in the properties predicted, as well as the datasets on which these networks have been trained. Moreover, training set sizes have been fairly limited, given that property labels, particularly from experimental methods, are scarce among the entirety of known chemical space. For example, the QM9 dataset[27-28] has 108k entries, and the ZINC dataset[29] was sampled to 250k entries in Gomez-Bombarelli et al.[23]

Here, we introduce an advanced deep learning approach, called DarkChem, that builds upon previous VAE work by incorporating several innovations and focuses on predicting chemical properties for use in metabolomics and non-targeted small molecule identification. We initially demonstrate DarkChem for (i) property prediction to create a massive *in silico* library, (ii) an initial small molecule identification test application, and (iii) example novel molecule generation, all focused on *m/z* (obtained from mass spectrometry after ionization) and collision cross section (CCS; obtained from ion mobility spectrometry). These properties have been demonstrated, in concert, to build evidence for the presence of molecules in complex biological samples[12, 30-36]. The mass-to-charge ratio has a long history for use in compound identification, and is the core feature around which most identifications are anchored in current non-targeted small molecule identification pipelines[37]. CCS is a measure of an ionized molecule's effective interaction surface with a buffer gas from ion mobility spectroscopy separations. Importantly, both properties can be consistently and accurately measured experimentally[38-46], as well as predicted computationally[12, 47-50].

A critical feature of DarkChem is its use of a unique 3-stage transfer learning method that enables the network to learn fundamental molecular structure representation from tens-of-millions of molecules before subsequent optimization of the network to improve its ability to predict chemical properties. This is highly valuable, as experimental chemical property training sets are often too small to take advantage of large and complex deep learning networks without risk of overtraining (i.e., trivially memorizing all, or portions, of the training set and preventing generalizability of the predictions). Thus we can increase the training set size for molecular property predictors despite limited experimental data. Since *m/z* is trivially calculated from chemical formula/structure, we have access to ~53 million structure-*m/z* pairs, but without CCS, from PubChem[51]. Additionally, the *in silico* Chemical Library Engine (ISiCLE) was used to generate *in silico* CCS for ~600k compounds from the Human Metabolome Database (HMDB)[52], the Universal Natural Product Database (UNPD)[53], and the Distributed Structure-searchable Toxicity (DSSTox)[54] database. Finally, we curated a set of 756 experimentally validated CCS values (metabolomics.pnnl.gov) from in-house data and from the literature[55-61]. Through a cascade of transfer learning iterations, our network is able to learn as much as possible from each dataset, enabling success with progressively smaller datasets without overfitting. Through this training regime, DarkChem is able to predict CCS to an average error of 2.5%, which is sufficient for immediate use by the metabolomics community to help build evidence for the presence of molecules and downselect candidate structures for samples run on ion mobility-mass spectrometry instruments, as we demonstrate in a small test application of a series of synthetic complex samples. Furthermore, we highlight DarkChem's generative capacity, wherein novel molecular structures can be created to match a set of desired experimental properties.

## EXPERIMENTAL SECTION

**DarkChem Implementation.** DarkChem was written in Python (version 3.6)[62] and uses Keras[63] with Tensorflow[64] backend. Training was performed using Marianas, a cluster with Nvidia Tesla P100 (16 nm lithography, 3584 CUDA cores at 1.19 GHz, 16 GB HBM2 memory) GPUs, provided by Pacific Northwest

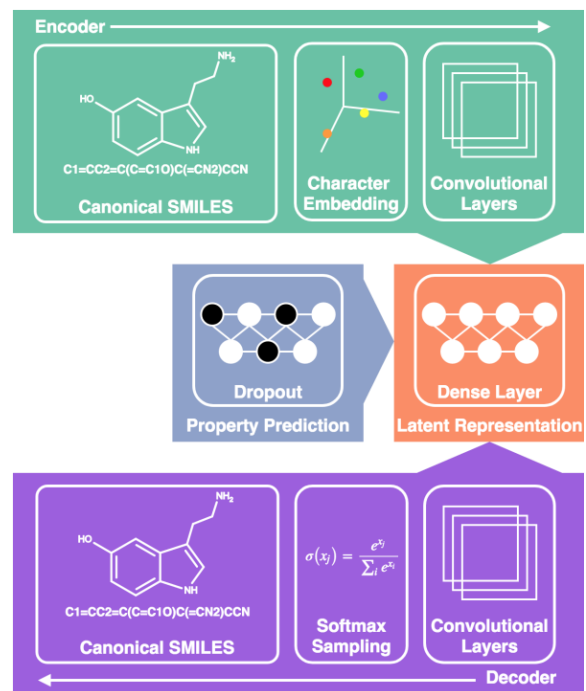

*Figure 1. DarkChem network schematic. The network involved an encoder (green), a latent representation (orange), and a decoder (purple). Additionally, a property predictor (slate) was attached to the latent representation. For the encoder, layers included SMILES input, character embedding, and a series of convolutional layers. The latent representation was a fully connected dense layer. The decoder was comprised of convolutional layers, followed by a linear layer with softmax activation to yield outputs. Finally, the property predictor was a single dense layer connected to the latent representation with 20% dropout.*

National Laboratory Research Computing. All code for the DarkChem architecture and supporting files are provided at github.com/pnnl/darkchem.

**Variational Autoencoder Architecture.** The overall DarkChem architecture consists of four components: 1) an encoder, consisting of a SMILES input encoder and convolutional layers, 2) a latent space, which holds the vector representation of molecular structure, 3) a decoder, consisting of convolutional layers and a SMILES character decoding layer, and 4) a property prediction layer. Components 1-3 comprise the VAE, which predicts inputs after encoding to a continuous numerical representation, and component 4 additionally predicts desired chemical properties, here accurate mass and CCS. Figure 1 shows a high-level schematic of the network architecture.

The network used for autoencoding SMILES input was structured similar to the VAE[25] introduced in Gomez-Bombarelli et al.[23], but with several key departures. The character set used involved 38 unique alphanumeric, punctuation, and symbol characters (e.g., 'C', '1', '(', '=') representing all characters present in the datasets used, plus a "pad" character (see Supporting Information, SI, Methods section). Datasets were downselected to molecules containing only carbon, hydrogen, nitrogen, oxygen, phosphorus, and/or sulfur (CHNOPS) atoms, and SMILES string lengths of 100 characters or fewer. This downselection was motivated by the application area (small molecule identification and metabolomics), wherein structures of interest are limited to CHNOPS molecules with low molecular weight (SMILES length serves as a surrogate filter for mass, as well as to limit network input size). SMILES strings less than 100 characters were extended to 100 characters with the pad character.

Each character was mapped to an arbitrary, but consistent, index, realizing a vector representation of inputs, which are passed to a 32-dimensional character embedding layer. This enables the network to learn a rich representation of the character set, rather than operate on arbitrarily assigned indices. Because of this step, vector inputs are evaluated against one-hot categorical encodings, as embedding layers cast integer indices as dense vectors for use in subsequent layers of the network. Thus, although an autoencoder, DarkChem's inputs (index vectors) and labels (one-hot encodings) differ in their representation, but only superficially.

Three convolutional layers with [9, 9, 10] filters and kernel size [10, 10, 11], respectively, follow the character embedding, each with rectified linear unit (ReLU) activation[65]. These connect to a linearly activated dense layer of 128 dimensions, corresponding to the latent vector representation of molecular structure. The variational components of the autoencoder are also initialized at this step as linearly activated dense layers, representing the mean and variance of the variational noise added to the latent representation. A Kullback-Leibler divergence term[66] (Equation 1) was added to the objective function evaluation in order to penalize departures from a mean of 0 and a variance of 1, ensuring normally distributed noise was added to the latent representation during training, scaled by hyperparameter epsilon (ε=0.8). Right side terms are the Kullback-Leibler divergence, $D_{KL}$; expected and observed probabilities $q_\phi$ and $p_\phi$, respectively, over a set of observed variables, x, and a set of latent variables, z, with joint distribution p(z, x). Left side terms are the number of samples, N; the standard deviation of the distribution, σ; and the mean of the distrubtion, μ.

$$D_{KL}(q_\phi(z|x)||p_\theta(z)) = -\frac{1}{N}\sum_{i=0}^{N} 1 + \log(\sigma) - \frac{\mu^2}{2} - \sigma$$

Eq. 1

The decoder connects directly to the latent dense layer and consists of three convolutional[67] ReLU layers with [9, 9, 10] filters and kernel size [10, 10, 11], respectively, as in the encoder portion of the network. Finally, a *softmax*-activated[68] dense layer, reshaped to match the dimensionality of the one-hot encoded targets, was added to predict final character sequences. The *softmax* outputs were evaluated using categorical cross entropy[69] (Equation 2) during training, but final outputs were decided using a beam search[70] decoder, an algorithm that yields the *k* most-probable discrete string predictions from the *softmax* outputs produced by the network. The network was optimized by AMSGrad[71] with default parameters except decay, which was set to 1E-8. Batch size during training was 32

**Property Prediction.** For multitask configurations in which labels are supplied for a semi-supervised training approach, the network additionally initializes a property prediction subnetwork that connects directly to the latent dense layer, but with 20% dropout such that property concepts are learned redundantly in the latent representation, with the intent of minimizing excess nonlinearity and overfitting[72]. A single, linearly activated dense layer with shape equal to the number of predicted labels (arbitrary, but in this work was of dimension two: CCS and *m/z*) is then used for property prediction.

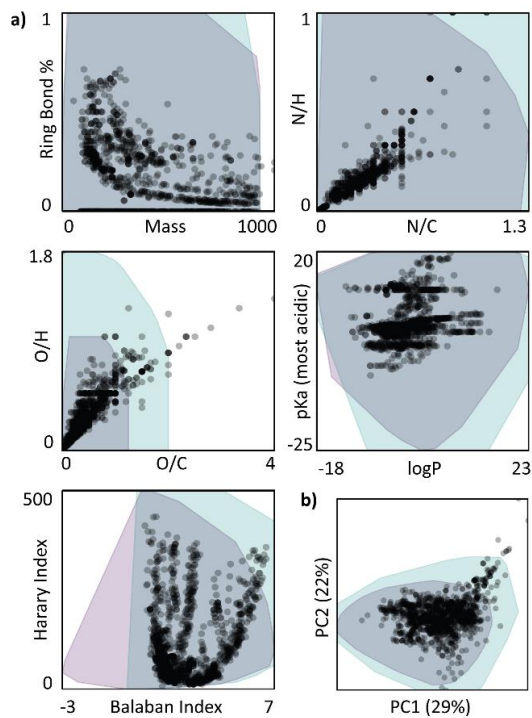

*Figure 2. Training set chemical space coverage. (a) Distribution of predicted properties. (b) Principal component analysis performed on the properties plotted in (a), with properties normalized to have a mean of 0 and standard deviation of 1. Purple is the convex hull for the PubChem dataset, blue is the convex hull for the* in silico *dataset, and black represents the experimental dataset. All convex hulls cover 99.5% of the underlying data (see Figure S1 for unfiltered plots).*

It is worth noting that CCS varies among multiple ion forms, or adducts, of a single parent molecule. Based on the ISiCLE and experimental training data sets we had available for this work, these include protonated, [M+H]$^+$, deprotonated, [M-H]$^-$, and sodiated [M+Na]$^+$ adducts, though there are many more possible adduct types. Separate networks were trained to predict CCS and *m/z* for each adduct type, but we will refer generally to CCS in reference to [M+H]$^+$, unless otherwise specified.

**Objective Function.** DarkChem is trained via a custom objective function that minimizes categorical cross entropy (Equation 2) between *softmax*-activated predictions and one-hot-encoded targets, where N represents the number of observations, J the number of classes (possible characters), and y and ŷ, the observed and expected variables, respectively. Additionally, a Kullback-Leibler (KL) divergence term (Equation 1) was included in the objective function evaluation to ensure that normally distributed noise was added to the latent representation during training by penalizing departures from mean 0 and variance 1. Categorical cross entropy and KL-divergence terms were weighted equally (i.e. representations in Equations 1 and 2 were summed without scaling).

$$CCE = -\frac{1}{N}\sum_{i=0}^{N}\sum_{j=0}^{J} y_j \log(\hat{y}_j) + (1 - y_j)\log(1 - \hat{y}_j)$$

Eq. 2

When predicting labels under a semi-supervised multitask learning configuration, a separate objective function was used to evaluate property prediction loss as the mean absolute percent error between the predicted and target property vector. Thus, the VAE loss was represented by categorical cross entropy and KL-divergence losses, while the property prediction loss, present during multitask training, was simply the mean absolute percent error loss of the predicted property vector. The two loss terms were weighted equally.

**Training.** Three datasets were used for training: PubChem[51]; the union of the Human Metabolome Database (HMDB)[52], the Universal Natural Products Database (UNPD)[53], and the Distributed Structure-Searchable Toxicity (DSSTox)[54] database with *in silico* predicted CCS values, henceforth the "*in silico* dataset"; and a curated library of molecules with experimental CCS values (*metabolomics.pnnl.gov*), which span a representative subset of known chemical space (Figure 2, Figure S1). The PubChem dataset was used to pretrain the VAE on a large set of SMILES strings (N=53,335,670) with calculated *m/z*. For the *in silico* dataset, along with SMILES and *m/z*, had associated CCS values, calculated using ISiCLE[12]. Thus, the *in silico* dataset is a larger (N=608,691) proxy to actual experimental CCS values (N=403, 486, and 371 for $[M+H]^+$, $[M-H]^-$, and $[M+Na]^+$ adducts, respectively; a combined 756 unique parent molecules), more amenable to training a large neural network.

We evaluated a number of training configurations in order to achieve success with progressively smaller datasets without overfitting. These included training directly on the small experimental dataset, training on *in silico* data only, and transfer learning configurations wherein the network is pretrained on PubChem and/or *in silico* data and subsequently "tuned" on experimental data. Transfer learning configurations also included pretraining with VAE-only and multitask (VAE plus property) networks. Additionally, in an effort to minimize overfitting effects, particularly with tuning on the small experimental dataset, we explored transfer learning configurations wherein the VAE weights were frozen, meaning only property predictor weights could vary during training with subsequent datasets. This effectively "freezes" the latent representation of molecular structure for subsequent training steps with smaller datasets. A summary of training configurations is depicted in Table S1, but this manuscript will focus on the network that involved: (i) train VAE and *m/z* predictor on PubChem, (ii) continue training on the *in silico* dataset, with the addition of CCS prediction, (iii) finish training the *m/z* and CCS predictor on experimental data, with frozen VAE weights. A schematic of the training paradigm is depicted in Figure 3.

In all training cases, data were shuffled during each epoch, and training was performed for 10,000 epochs with an early stop callback (patience 1,000) to avoid overfitting. Validation was performed on a random 10% subset for PubChem and *in silico* dataset training. For the experimental dataset, 100 iterations of repeated random subsampling validation with 10% holdout were performed. Select learning curves are depicted in Figures S2 and S3.

**Hyperparameter Selection.** The instantiation of the network detailed here contains specific selections for all hyperparameters, but the network is architected such that all parameters are configurable through the command line. This includes character embedding dimension; number of filters, kernel sizes, and number of convolutional layers; latent dimension size; epsilon, which scales the noise added during training; and dropout fraction on the latent vector for property prediction. Additionally, several aspects of the network architecture are detected automatically, including length of input vectors, number of unique characters, and number of target labels (for multitask training). Using this generalized framework, a sweep over selected parameters, including latent dimension size, number of filters, kernel size, noise parameter epsilon, dropout, and embedding dimension, wherein each parameter was varied one at a time, was performed. Though not exhaustive, this cursory evaluation led to a reasonably performing network, successful for this application.

***In silico* CCS Library Generation.** CCS values were determined from SMILES found in the PubChem and *in silico* datasets (i.e., HMDB, UNPD, and DSSTox) through the trained DarkChem network to generate CCS for $[M+H]^+$, $[M-H]^-$, and $[M+Na]^+$ adducts. This was done without adding the normally distributed noise (epsilon) that was added to the latent representation during training. Membership of each CCS value (N = 161,965,516) was assessed whether they were inside or outside of the same chemical space as the experimental training set. This was performed by evaluating membership within the convex hull encompassing the training set in the first eight dimensions from the principal component analysis (PCA)[73] of DarkChem's latent space (see Figure S4 for explained variance by dimension). Those found within the chemical space were included as entries into the final *in silico* CCS library (N = 90,995,413).

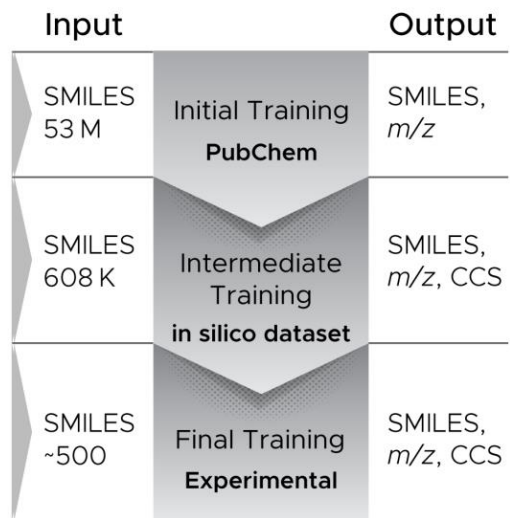

*Figure 3. Training schematic. DarkChem was initially trained on ~53 million inputs from PubChem, wherein m/z was the only predicted property. Weights from this network were used to seed the next, which involved training on the ~600,000 in silico dataset, with m/z and in silico CCS labels. The further trained weights seeded the final training step, which involved ~500 inputs with m/z and experimental CCS. For some network configurations, weights were frozen (i.e. no longer updated) to prevent overfitting to smaller datasets, in particular the experimental dataset. The various training configurations investigated can be seen in Table S1.*

**Beam Search Decoder.** Although not used during training, we have additionally implemented a beam search decoder to realize *k* discrete strings from *softmax* predictions, where *k* is the beam width, yielding the *k* most probable SMILES sequences. Thus, beam search may be used for all generative applications, offering several advantages over the *argmax* operator necessitated during training.

**Generative Mode.** When using the network in a generative capacity, the desired outcome involves predicting candidate structures from known property signatures (e.g. CCS, *m/z*) obtained from experimental instruments (e.g. ion mobility spectroscopy, mass spectroscopy). The training paradigm used enables effective entanglement without excessive nonlinear overfitting such that PCA[73] is able to project the latent representation into a space that correlates, in at least one dimension, with desired properties (see Results and Discussion). Thus, one can start with a molecule of a certain *m/z* and CCS and move orthogonally to the respective correlated PCA dimension(s) to yield putative structures with shared property information.

PCA was performed using *scikit-learn*[74-76], and correlation with desired properties was evaluated using the correlation coefficient between each principal component and each property. Latent shaping was considered successful when (i) at least one principal component correlated heavily with predicted properties, and (ii) at least one principal component was invariant to (or uncorrelated with) predicted properties. With (i) and (ii) satisfied, putative structures were generated by moving in the dimension(s) defined by (ii) and subsequently performing the inverse transform on the PCA vector to yield a latent vector representation. Resulting latent vectors were decoded using beam search and additionally checked to ensure they mapped to valid SMILES strings using *rdkit*[77].

**Synthetic complex samples and analytical experiments:** Samples were provided through the U.S. Environmental Protection Agency - Non-Targeted Analysis Collaborative Trial (ENTACT) challenge, a blinded inter-laboratory challenge[31], designed for the objective testing of non-targeted analytical chemistry methods using a consistent set of synthetic mixtures. Each mixture contained between 95 and 365 compounds, all selected from the EPA ToxCast chemical library. Further details on ENTACT are outlined in Sobus et al.[78] and Ulrich et al.[79] The ten synthetic mixtures and blanks were analyzed using a drift tube ion mobility spectrometry-mass spectrometer[58, 80] and a 21-Tesla Fourier transform-ion cyclotron resonance spectrometer-mass spectrometer (FTICR-MS)[81-83] in both positive (+) and negative (-) ionization modes. Additional experimental details are provided in Nuñez et al.[31] Evidence for the presence of molecules in each sample was assessed using the Multi-Attribute Matching Engine (MAME), a modular Python package that performs feature downselection and a weighted scoring system that, in the case of the ENTACT study, was used to assign compounds as suspected present if their score surpassed a defined threshold. Any compounds labeled as suspected present that, after unblinding, were found to be intentionally spiked in are considered as true positives.

**RESULTS AND DISCUSSION**

Motivation for the VAE network configuration was multifaceted: (i) VAEs are useful as property prediction frameworks, (ii) we hypothesized that a VAE could improve upon existing methods – both first-principles simulation and other machine-learning based approaches – in terms of accuracy and throughput, and (iii) VAEs can be used for use in a generative capacity. That is, a VAE learns a continuous numerical, or latent, representation of molecular structure (and associated properties) such that novel candidate structures with desired properties can be generated for use in untargeted metabolomics and small molecule identification applications. The multitask training configuration was designed to coax the network into encoding molecular properties explicitly, despite being emergent properties of structure, without supplying this information directly (i.e. encoding through prediction rather than via input). Thus, results are interpreted in both capacities: the network as a property predictor *and* the network as a generative tool for small molecule identification and discovery. Additionally, the value added by performing this training simultaneously is assessed, as we demonstrate synergistic effects of combining an autoencoder with property prediction.

**Reconstruction Accuracy.** Although not explicit in the objective loss, reconstruction accuracy, the mean per-character absolute difference between input and predicted SMILES sequences, was used as an intuitive performance assessment. The network trained on the limited (N=403 for [M+H]$^+$ adducts) experimental data only yielded validation reconstruction accuracy of 78.5%. This is in contrast to the transfer learning (final production-mode) network, which achieved validation reconstruction accuracy of 98.9% for the experimental dataset and 99.0% for the *in silico* dataset, indicating that a sizeable and varied dataset was required to learn a general representation of chemical structure. Out-of-sample validation (network trained on experimental values only, evaluated with *in silico* data) further confirmed this discrepancy, as reconstruction accuracy was only 70.8% with out-of-sample data. Thus, we confirmed the power of our 3-stage transfer learning method, taking advantage of much larger training sets than is typically possible, compared to traditional single stage learning approaches.

It is worth noting that reconstruction accuracy, though integral to the success of training a VAE, is only a proxy for the true objective of the network. Reconstruction accuracy represents the network's ability to recreate an input SMILES string from its associated latent representation, despite added noise. The added noise ensures the latent space is continuous, rather than discrete per each entry in the dataset, but at what point should a noise perturbation yield a new structure? Moreover, during training, if the added noise does yield a new structure, the network is penalized as said structure does not match the input. This is antithetical to the goal of the VAE, as it functions to generate new structures from a given input following perturbation, yet is penalized during training when this occurs. When considering the network in a generative capacity, adding noise to a known latent vector should, with sufficient noise magnitude, yield a new, valid

SMILES structure, not the input. But the objective function is unable to reflect this without significant modification. Still, training a VAE, which attempts to faithfully recreate inputs despite added noise, functions as a reasonable proxy to a valid SMILES discriminator, as evidenced by the ability of the networks trained on *in silico* data to generalize to out-of-sample experimental data, with and without experimental fine tuning.

**Property Prediction.** Key to the success of this work was the use of a shared latent space. That is, a latent space that simultaneously encodes a continuous numerical representation of structure and associated chemical properties. Coupled with the use of a relatively small (with respect to number of layers) property decoder, which forces the latent space to encode this chemical property information, the resulting latent representation learned a rich representation.

In most cases, networks were able to achieve reasonable success when predicting in-sample CCS and *m/z*. Training on experimental data only, validation error was 3.5% and 2.2% for CCS and *m/z*, respectively. The best performing network in terms of CCS prediction achieved CCS and *m/z* errors of 2.5% and 0.7%, respectively. The final transfer learning configuration, selected for its advantages in generality and latent space correlations, had validation error of 3.0% and 0.4% for CCS and *m/z*, respectively. A summary of property prediction errors for evaluated training configurations can be found in the Supporting Information (Table S1). Although we focus on CCS for [M+H]$^+$ adducts, networks were additionally trained to predict [M-H]$^-$ and [M+Na]$^+$, each with comparable reconstruction accuracy (99.3% and 99.5%, respectively), *m/z* prediction error (0.4% and 0.3%, respectively), and CCS error (3.1% and 2.5%, respectively).

The network's capacity to predict properties directly from chemical structures (as represented by canonical SMILES strings) represents a new tool for the metabolomics and small molecule identification community, particularly concerning the prediction of CCS (*m/z* is important for using the network in a generative capacity, but this property is trivial to calculate otherwise). Previous efforts have been able to achieve 3.2% error using first-principles simulation[12] and 3.3% error via property-based machine learning approaches[1-2], and 3.0% error via a non-generative, SMILES-based deep-learning approach[18], each evaluated on the experimental data. The method detailed here uses structure, represented by SMILES string, to predict properties directly, and is able to do so with lower CCS error for most adducts. Additionally, prediction time (after training) is orders of magnitude faster than first-principles simulation (milliseconds on a laptop compared to node-hours on a high performance computer)[12], and does not require chemical property calculation needed for use with property-based methods, such as MetCCS[2]. Finally, DarkChem is a generative approach, enabling usefulness beyond just property prediction. With consideration to accuracy and computational efficiency of this method, it emerges as a highly useful tool for *in silico* chemical property library expansion for applications in standards-free small molecule identification and metabolomics.

**Property Correlation.** The property "concepts" learned by the network through supervised prediction were evaluated in terms of how select dimensions of the latent representation were

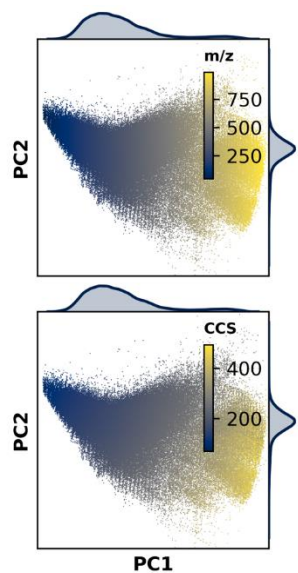

*Figure 4. Latent space. The first two principal components of the 128-dimensional representation are shown, colored by predicted property value (top: m/z, bottom: CCS). The representation is a 2D binned statistic of the mean, with grid size 384 in each principal component dimension. A kernel density estimator is also shown for each principal component dimension, emphasizing density of the distribution. Clear correlations to m/z and CCS are observed, largely across the first principal component (see Figure S6 for correlation plots).*

correlated – and uncorrelated – with *m/z* and CCS (see Figure S5 for latent variable distributions). Correlation analysis was also performed in PCA space. Properties were plotted against the most and least correlated latent dimensions, as well as the most correlated PCA dimension, in Figure S6. Key to this analysis was the fact that dimensions, to some degree, specialize in human-interpretable information (i.e., the prediction of chemical properties), as indicated by different latent dimensions correlating most heavily with *m/z* and CCS, respectively, as well as multiple dimensions exhibiting no correlation with predicted properties, presumably specializing in other network concepts. Further elucidation of human-interpretable network concepts learned during training is a target for future effort.

Additionally, the first principal component exhibited even greater correlation with *m/z* and CCS than any individual latent dimension (Figure 4). Thus, moving along those remaining principal components uncorrelated with *m/z* and CCS proved useful in a generative capacity for which putative structures could be yielded for a given *m/z* and CCS. Traversing dimensions invariant to *m/z* and CCS enables generation of known and potentially novel candidates that can be matched to currently annotatable – due to lack of authentic reference values – experimental signals.

**Training Paradigm.** Although only a select few of the networks evaluated in this work (see Table S1) were useful for generative applications and/or property prediction, the poorly performing networks revealed several interesting insights. Reconstruction accuracy was low when training on experimental values directly, thus necessitating use of the *in silico* dataset and/or the PubChem dataset, as each were of sufficient size to yield satisfactory reconstruction accuracy and property prediction error, if applicable. However, as evidenced by the high property prediction error in networks seeded with the frozen autoencoder weights of the PubChem-trained networks (N4a, N5a, N5b), high reconstruction accuracy did not indicate a representation of molecular structure sufficient for *m/z* and CCS prediction. Thus, the intermediate step of training on *in silico* data allowed property concepts to form in the latent representation, and also enabled the weights of the autoencoder portion to be frozen during training on the small experimental dataset to avoid overfitting. Although it was possible to achieve high reconstruction accuracy and low property prediction error training on just the *in silico* dataset followed by the experimental dataset, the learning configurations that included PubChem were preferred to include a larger number of varied training examples, particularly considering that the experimental data is completely subsumed by the *in silico* dataset, but PubChem contains molecules outside the *in silico* dataset (based on convex hull analysis, see Figure S7). This suggests that networks trained with PubChem data would generalize favorably to molecules in this region, compared to those trained without.

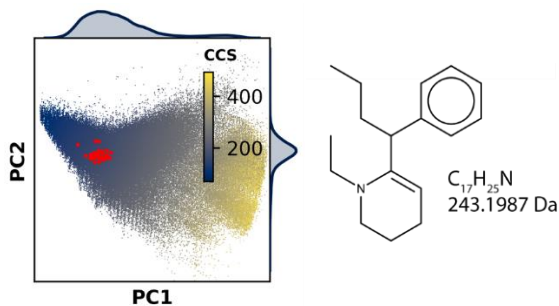

*Figure 5. Phencyclidine analogue. By seeding latent space with a known set of NMDA receptor PCP site antagonists (a-b), a large number of putative phencyclidine (PCP) analogues were yielded. Of these, a novel analogous structure, 3-{8,9-dihydro-5H-benzo[7]annulen-1-yl}-2-propylazetidine, was found with 5 ppm mass error and 0.2% error in predicted CCS (experimental CCS not evaluated).*

**Chemical Space Coverage.** Given the potentially rich representation of chemical structure encoded in each latent vector, categorizations in terms of dataset and chemical class were performed in principal component space for visual interpretation. For dataset source, convex hulls were constructed for each of PubChem, HMDB, UNPD, and DSSTox, as well as the convex hull of their union, and plotted in Figure S7. Datasets largely overlapped, but some spanned distinct regions of the PCA representation. Notably, the HMDB, which was the only dataset containing a high number of lipids – structures with high *m/z* and CCS – was also the only dataset to occupy the rightmost portion of the PCA convex hull. Similarly, PubChem was the only dataset with a non-biological focus; it thus spanned the largest portion of the PCA representation of latent space, particularly unique in its coverage of the left- and upper- most portions. Similar analysis was performed for chemical class (defined by ClassyFire[84]), as depicted in Figure S8. Hull separations were distinctly visible for several classes, while others depicted regions of significant overlap, indicating the latent representation encoded, in at least some capacity, a distinction among molecules from human-assigned ontology. Methods for hull analysis are detailed in the Supporting Information, Methods section.

***In Silico* Library**. DarkChem was used to generate CCS predictions for a set of 3 adducts for molecules from PubChem, HMDB, UNPD, and DSSTox. CCS values for [M+H]+, [M-H]−, and [M+Na]+ adducts are made available in the SI (and will be kept updated at metabolomics.pnnl.gov). To ensure conservative predictions, that is, only predicting values for molecules similar to those in the experimental training set, a convex hull of the experimental data was constructed from their associated latent vectors. Compounds from PubChem, HMDB, UNPD, and DSSTox that fell within the convex hull of experimental values were used to build the library, which currently contains 90,995,413 entries, and is being updated as more data becomes available. The initial library is provided in the SI, with the most current version of the library being available at metabolomics.pnnl.gov.

**Analysis of synthetic complex samples.** In our initial study of the ENTACT challenge, we found evidence for 618 true positive compounds that we suspected were present from our analysis using MAME. In that study, calculated CCS from ISiCLE increased the confidence of 84% of molecules that were correctly determined to be present in the samples, showcasing its importance in as an additional property to mass and isotopic signatures. Compared to the true positive experimental standards spiked in these samples that were uniquely identified, calculated CCS errors for DarkChem values was 2.8% and 2.6%, for those CCS that fell within the same latent space as the experimental training set (N = 37), or outside (N = 25), respectively. This is comparable to 3.2% error when using Standard ISiCLE CCS values, as were originally used in the study, and a 2.9% error when using DeepCCS. This out-of-sample test demonstrates consistent CCS error values compared to the initial validation set.

**Generative Modes.** The network resulting from the cascade of transfer learning iterations was used in two generative applications: first, an interpolation between adenine and cholesterol (Figure S9) and second, generation of a putative compound analogous to a set of known PCP analogues, with a specifically targeted *m/z* and CCS value (Figure 5).

For interpolation, a direct linear interpolation – that is, projecting a vector from the latent representation of molecule A to the latent representation of molecule B and sampling along its length – caused sampling of empty regions of latent space, meaning interpolated latent vectors decoded to invalid SMILES strings in some cases. To ameliorate this phenomenon, the closest training example to each interpolated point along the interpolation vector was used to seed a number of putative structures. From these sets, molecules were selected to minimize the standard deviation of latent space distance between each interpolate. This was in an attempt to produce a set with as-smooth-as-possible transitions. These empty regions of latent space represent a shortcoming of the network, which will be addressed in future efforts. A demonstrative interpolation between adenine and cholesterol is shown in Figure S9.

For analogue generation, an initial set of known N-methyl-D-aspartate (NMDA) receptor PCP site antagonists was used to seed a subregion of latent space. The mean and standard deviation of the latent representations of these known antagonists were used to sample a normal distribution to yield putative analogues. The putative list was filtered by *m/z* and CCS error to find candidates closely resembling PCP in their property signature. The most similar novel structure is shown in Figure 5, with *m/z* error of 5 ppm (calculated from formula), and predicted CCS error of 0.2%, as well as the clustering of the known NMDA receptor antagonists in latent space (compressed to two dimensions by PCA).

## CONCLUSION

This article introduces DarkChem, a framework for the characterization of small molecules that can be used for putative identifications in complex mixtures directly from experimental signals, such as *m/z* from mass spectrometry and CCS from ion mobility spectrometry. DarkChem offers a number of advancements over previous works in that 1) properties are predicted directly from structure, as opposed to calculated chemical properties or other derived features, 2) predicted properties are relevant to the field of metabolomics, particularly for applications involving putative identifications using untargeted IMS/MS pipelines, and 3) the network was trained on the largest dataset to-date, improving learned molecular concepts and property predictions with each successive dataset (PubChem, *in silico*, experimental). Combined, these advances position DarkChem as a highly useful offering in the metabolomics community and beyond, particularly considering that the framework supports training with arbitrary properties. That is, in addition to, or instead of, *m/z* and CCS, to meet the requirements of putative identifications from experimental data acquisitions involving varying instrument arrays.

## ASSOCIATED CONTENT

**Supporting Information.** The Supporting Information is available free of charge on the ACS Publications website.
- SI Methods and SI Figures
- DarkChem source code and network weights
- DarkChem CCS library

## AUTHOR INFORMATION

**Corresponding Author.**
*E-mail: ryan.renslow@pnnl.gov
**Author Contributions.** The manuscript was written through contributions of all authors. All authors have given approval to the final version of the manuscript.

## ACKNOWLEDGEMENTS

This research was supported by the Pacific Northwest National Laboratory (PNNL) Laboratory Directed Research and Development program via the Deep Science Agile Initiative; and the National Institutes of Health, National Institute of Environmental Health Sciences grant U2CES030170. PNNL is operated for DOE by Battelle Memorial Institute under contract DE-AC05-76RL01830. The authors thank Dr. Thomas Metz (PNNL) for valuable discussion and comments on the manuscript.